\documentclass{PoS}
\usepackage{graphicx}
\usepackage{amssymb}
\usepackage{amsmath}
\usepackage{multirow}
\usepackage{dcolumn}
\usepackage{nicefrac}

\title{Delta and Omega electromagnetic form factors in a covariant three-body approach}

\ShortTitle{$\Delta$ and $\Omega$ electromagnetic form factors in a covariant 
three-body approach}

\author{\speaker{H\`elios Sanchis Alepuz}\\
        Justus-Liebig-Universit\"at Giessen\\
        E-mail: \email{helios.sanchis-alepuz@theo.physik.uni-giessen.de}}

\author{Reinhard Alkofer\\
        Institut f\"ur Physik,
Karl-Franzens-Universit\"at,
Universit\"atsplatz 5,
A-8010 Graz, Austria\\
        E-mail: \email{reinhard.alkofer@uni-graz.at}}

\author{Richard Williams\\
        Institut f\"ur Physik,
Karl-Franzens-Universit\"at,
Universit\"atsplatz 5,
A-8010 Graz, Austria\\
        E-mail: \email{richard.williams@uni-graz.at}}

\abstract{We study baryons as three-body systems using the QCD degrees of freedom 
in the framework of covariant Bethe-Salpeter equations. 
The interaction among quarks is reduced to a 
vector-vector interaction via a single dressed-gluon exchange (Rainbow-Ladder 
truncation). The formalism allows for the study of the hadron spectrum as well as 
their internal properties. We will present the calculation of electromagnetic 
properties of spin-3/2 baryons. The model independent features of our results are 
assessed using two different models for the dressings.}

\FullConference{Xth Quark Confinement and the Hadron Spectrum\\
                 8 - 12 October 2012\\
                 TUM Campus Garching, Munich, Germany}

\begin{document}

\section{Introduction}

Whereas the internal structure of the nucleon is rather well established experimentally,
this is not the case for the next baryonic state, the $\Delta(1232)$. The reason is its
unstable nature which makes the study of its properties a challenging task. Insight into
the $\Delta$'s internal structure can be gained by studying its electromagnetic
properties. 

Since the $\Delta$ is a spin-$\nicefrac{3}{2}$ particle, its electromagnetic structure
is characterized by four electromagnetic form factors; these are the electric monopole
$G_{E0}$, electric quadrupole $G_{E2}$, magnetic dipole $G_{M1}$ and magnetic octupole
$G_{M3}$. They are functions of the photon momentum $Q$, their respective values at
$Q=0$ are the static electromagnetic moments of the baryon. Experimentally, only the
magnetic dipole moments (and, of course, the electric charge) are known for the
$\Delta^+$ and $\Delta^{++}$, albeit with large errors. Indirectly, the study of the
$\Delta\to N\gamma$ decay indicates a deviation of the $\Delta$ from sphericity and
therefore a non-vanishing value for $G_{E2}$.

We present here a calculation of the electromagnetic moments using a covariant
Bethe-Salpeter approach. This approach provides a unified quantum-field-theoretical
description of meson and baryon properties (see {\it e.g.}, 
Ref.~\cite{Eichmann:2009zx} and
references therein). With the flavor-blind interaction kernels we currently use, the
$\Omega(1672)$ is identical to the $\Delta$ but calculated at higher quark mass.
Therefore we present also its electromagnetic moments. A more complete study of the
$\Delta$ and $\Omega$ electromagnetic properties at non-vanishing momentum transfer 
is presented in \cite{FFpaper}.

\section{Framework}

In this section we briefly describe the calculation of baryon properties using a
covariant three-body Bethe-Salpeter approach; for details we refer to 
Refs.~\cite{FFpaper,Eichmann:2009qa,Eichmann:2011vu,SanchisAlepuz:2011jn} 
and references therein.

\begin{figure*}[ht!]
 \begin{center}
  \includegraphics[width=\textwidth,clip]{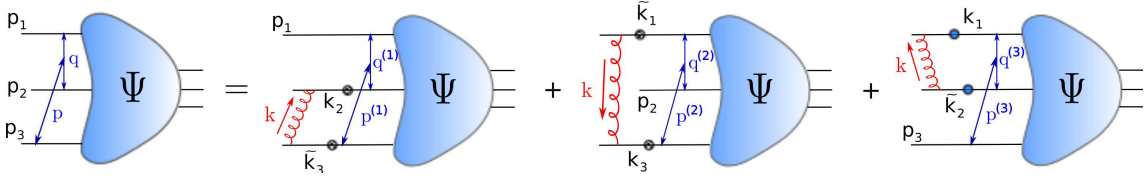}
 \end{center}
 \caption{Covariant Faddeev equation in the
Rainbow-Ladder truncation.}\label{fig:FaddeevRLeq}
\end{figure*}

All the information about the baryon structure is encoded in the Faddeev amplitudes
$\Psi$. They are obtained by solving the covariant Faddeev equation
\begin{flalign}\label{eq:faddeev_eq}
\Psi_{\alpha\beta\gamma\mathcal{I}}(p,q,P) ={}&\int_k  \left[
\widetilde{K}_{\beta\beta'\gamma\gamma'}(k)~S_{\beta'\beta''}(k_2)
S_{\gamma'\gamma''}(\tilde{k}_3)~
\Psi_{\alpha\beta''\gamma''\mathcal{I}}(p^{(1)},q^{(1)},P)\right.\nonumber\\
&\quad \left.
+\widetilde{K}_{\alpha\alpha'\gamma\gamma'}(-k)~S_{\gamma'\gamma''}(k_3)
S_{\alpha'\alpha''} (\tilde{k}_1)~
\Psi_{\alpha''\beta\gamma''\mathcal{I}}(p^{(2)},q^{(2)},P)
\right. \nonumber\\
&\quad  \left. +
\widetilde{K}_{\alpha\alpha'\beta\beta'}(k)~S_{\alpha'\alpha''}(k_1)
S_{\beta'\beta''}(\tilde{k}_2)~
\Psi_{\alpha''\beta''\gamma\mathcal{I}}(p^{(3)},q^{(3)},P)\right] ~,
\end{flalign}
with $p$ and $q$ being the relative momenta, as depicted in Fig. \ref{fig:FaddeevRLeq} 
and defined in \cite{FFpaper}, and $P$ is the total momentum. $S_{\alpha'\alpha''}(k_1)$
is the fully dressed quark propagator.
Here we use a Ladder truncation for the two-body interaction kernel
\begin{equation}
	\widetilde{K}= 4\pi \,Z_2^2 \,\frac{\alpha_{\rm eff}(k^2)}{k^2}\,
	T_{\mu\nu}(k)\,\gamma^\mu \otimes \gamma^\nu\,\,
	\label{eqn:ladder}
\end{equation}
with $Z_2$ being the quark renormalization constant, $T_{\mu\nu}(k)$ the $k$-transverse
projector and $k$ the gluon momentum. In addition we neglect three-body irreducible
interactions.

Eq.\ (\ref{eq:faddeev_eq}) requires the fully dressed quark propagator and the effective
interaction $\alpha_{\rm eff}$ in the two-body interaction kernel as input. For
consistency, the quark propagators are obtained by solving the
quark Dyson-Schwinger equation (DSE) in the Rainbow truncation such that also for these 
equations $\alpha_{\rm eff}$ is the only needed input. 

\begin{figure}[hbtp]
 \begin{center}
  \includegraphics[width=0.7\textwidth,angle=0,clip]{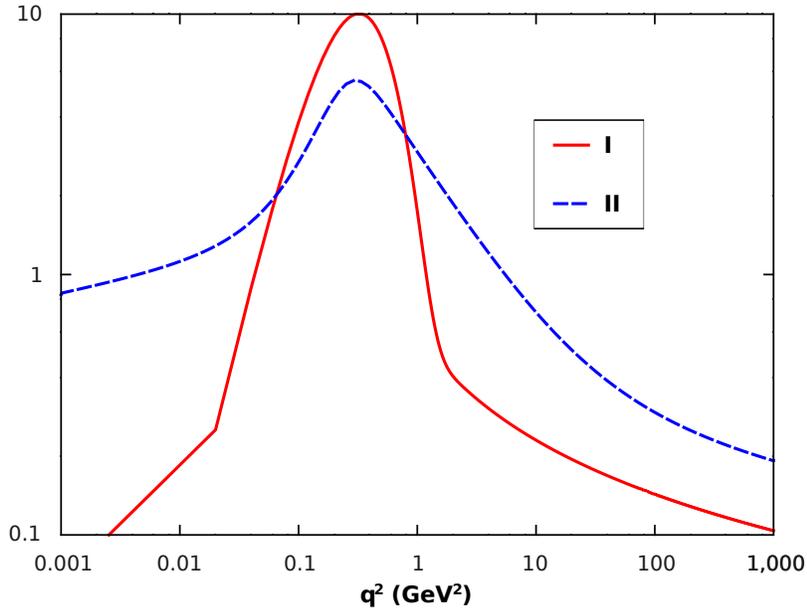}
 \end{center}
 \caption{Comparison of the behavior of $\alpha_{\rm eff}(q^2)$ for model \textbf{I} 
 \cite{nucl-th/9708029,nucl-th/9905056} and model \textbf{II} \cite{arXiv:0804.3478}
 used for the 
 effective interaction.}\label{fig:Int_comparison}
\end{figure}

The Rainbow-Ladder truncation thus relies on the choice of an effective interaction
$\alpha_{\rm eff}$. To learn about the model dependence of our results we compare two
different \textit{ans\"atze}: the first one is the Maris-Tandy model
\cite{nucl-th/9708029,nucl-th/9905056} which was originally invented to optimally 
reproduce light pseudoscalar observables; the second is a model designed to obtain the
$U_A(1)$ anomaly  enhanced $\eta'$-mass via the Kogut-Susskind mechanism
\cite{Kogut:1973ab,von Smekal:1997dq} and also to describe correctly pseudoscalar meson
properties \cite{arXiv:0804.3478}. Both models, which we denote as models \textbf{I} and
\textbf{II}, are compared in Figure \ref{fig:Int_comparison}; for a detailed description
of both of them we refer to  Refs.~\cite{FFpaper,SanchisAlepuz:2011jn}. 

\begin{figure*}[hbtp]
 \begin{center}
  \includegraphics[width=\textwidth,clip]{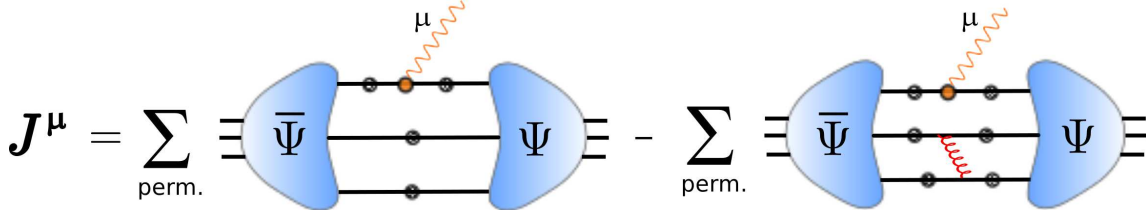}
 \end{center}
 \caption{Calculation of the baryon electromagnetic current in
the Rainbow-Ladder truncation.}\label{fig:FFeq_RL}
\end{figure*}

Note that Eq.\ (\ref{eq:faddeev_eq}) determines besides the Faddeev amplitudes also
the baryon mass: Only for the bound state mass $P^2=-M_B^2$ it will be fulfilled. 
In addition we want to remark that being constructed from the tensor product of three
Dirac spinors the number of independent tensor components of the Faddeev amplitudes
is quite large, 64 for the
nucleon and 128 for the $\Delta$ or $\Omega$.
Once the Faddeev amplitudes are obtained for given baryon quantum numbers 
further information  about the considered baryon can be extracted.
For example, one can use them to calculate the baryonic electromagnetic current using
the equation 
\begin{flalign}\label{eq:FFeqRL}
 J_{\mathcal{I}'\mathcal{I}}^\mu=\int_p\int_q\bar{\Psi}_{\beta'\alpha'\mathcal{I
}'\gamma'}(p_f^{\{1\}},q^{\{1\}}_f,P_f)\left[\left(S(p_1^f)\Gamma^\mu(p_1,
Q)S(p_1^i)\right)_{\alpha'\alpha}S_{\beta'\beta}(p_2)S_{\gamma'\gamma}
(p_3)\right]\times\nonumber\\
~~~~~~~~~~~~~~~~~~~~~~~~~~~~~~\left(\Psi_{\alpha\beta\gamma\mathcal{I}}(p^{\{1\}
}_i,q^{\{1\}}_i,P_i)-\Psi^{\{1\}}_{\alpha\beta\gamma\mathcal{I}}(p^{\{1\}}_i,q^{
\{1\}}_i,P_i)\right)\nonumber\\
+\int_p\int_q\bar{\Psi}_{\beta'\alpha'\mathcal{I}'\gamma'}(p^{\{2\}}_f,q^{\{2\}}
_f,P_f)\left[S_{\alpha'\alpha}(p_1)\left(S(p_2^f)\Gamma^\mu(p_2,
Q)S(p_2^i)\right)_{\beta'\beta}S_{\gamma'\gamma}(p_3)\right]\times\nonumber\\
~~~~~~~~~~~~~~~~~~~~~~~~~~~~~~\left(\Psi_{\alpha\beta\gamma\mathcal{I}}(p^{\{2\}
}_i,q^{\{2\}}_i,P_i)-\Psi^{\{2\}}_{\alpha\beta\gamma\mathcal{I}}(p^{\{2\}}_i,q^{
\{2\}}_i,P_i)\right)\nonumber\\
+\int_p\int_q\bar{\Psi}_{\beta'\alpha'\mathcal{I}'\gamma'}(p^{\{3\}}_f,q^{\{3\}}
_f,P_f)\left[S_{\alpha'\alpha}(p_1)S_{\beta'\beta}
(p_2)\left(S(p_3^f)\Gamma^\mu(p_3,Q)S(p_3^i)\right)_{\gamma'\gamma}\right]
\times\nonumber\\
~~~~~~~~~~~~~~~~~~~~~~~~~~~~~~\left(\Psi_{\alpha\beta\gamma\mathcal{I}}(p^{\{3\}
}_i,q^{\{3\}}_i,P_i)-\Psi^{\{3\}}_{\alpha\beta\gamma\mathcal{I}}(p^{\{3\}}_i,q^{
\{3\}}_i,P_i)\right)~,
\end{flalign}
which is depicted in Fig.~\ref{fig:FFeq_RL}. We refer again to Ref.\ 
\cite{FFpaper} for a definition of the kinematical variables. 
From the current $J$ one can extract the electromagnetic
form factors via the appropriate contractions (see, e.g. \cite{Nicmorus:2010sd}). The
non-perturbative quark-photon vertex required to calculate the electromagnetic current
is obtained by solving the vertex inhomogeneous Bethe-Salpeter equation in the
Rainbow-Ladder truncation.

\section{Results}

The electric monopole moment is not very interesting since it gives the electric 
charge of the baryon and it is fixed by the normalization of the Faddeev amplitudes. 
Nevertheless, related to its evolution with respect to the photon momentum is the 
electric charge radius
\begin{equation}
\langle r_{E0}^2 \rangle = -\frac{6}{G_{E0}(0)}\left.\frac{dG_{E0}}{dQ^2}\right|_{Q^2=0}~.
\end{equation}
We show our results for the $\Delta$ in Table \ref{tab:delta_moments} and for the 
$\Omega$ in Table \ref{tab:omega_moments} and compare to lattice results at different 
pion masses \cite{Alexandrou:2009hs,Alexandrou:2009nj,Alexandrou:2010jv}.  
One sees that there seems to be a significant disagreement between our results and 
the corresponding lattice data. However, a large 
part of this disagreement is simply due to the different values obtained for the  
decisive scale of the problem, namely, the baryon mass. 
This disagreement is, therefore, 
largely reduced by comparing the dimensionless quantity 
$\langle r_{E0}^2 \rangle M^2_\Delta$. 
In this case model \textbf{I} gives a result consistent with lattice data, whereas 
model \textbf{II} overestimates it somewhat.

\begin{table*}
\begin{center}
\renewcommand{\arraystretch}{1.6}
\begin{tabular}{l|cc|ccc|c|} 
~~~~~~~~~~~~~~~~~~~ ~~~~~~~~~~~
& F-\textbf{I} & F-\textbf{II} & DW1 & DW2 & DW3 & Exp.\\[2mm]
\hline\hline
$M_\Delta (\textnormal{GeV})$ & 1.22 & 1.22 & 1.395~(18) & 1.559~(19)
& 1.687~(15) & 1.232~(2)
\\
\hline
$\langle r_{E0}^2\rangle  (\textnormal{fm}^2)$ & 0.50 & 0.61 & 0.373~(21) &
0.353~(12) & 0.279~(6) &
\\
\hline
$\langle r_{E0}^2\rangle  M_{\Delta}^2$ & 0.75 & 0.91 & 0.726~(36) & 0.858~(25) &
0.794~(14) &
\\
\hline
$G_{M1}(0)$ & 2.38 & 2.77 & 2.35~(16) & 2.68~(13) & 2.589~(78) &
3.54$^{+4.59}_{-4.72}$ \\ \hline
$G_{E2}(0)$ & -0.67 & -0.99 & -0.87~(67) & -1.68~(88) & -0.71~(49) & \\ \hline
\end{tabular}
\caption{Comparison of static results for the $\Delta^{+}$. We compare our results 
for models \textbf{I} and \textbf{II} (F-\textbf{I} and F-\textbf{II}, respectively) with a lattice calculation with
dynamical Wilson fermions at $m_\pi=384$~MeV (DW1), $m_\pi=509$~MeV (DW2) and
$m_\pi=691$~MeV (DW3) \cite{Alexandrou:2009hs,Alexandrou:2009nj}. For
$G_{M1}(0)$ we also compare with the experimental value
\cite{Kotulla:2002cg,Beringer:1900zz}.}\label{tab:delta_moments}
\end{center}
\end{table*}

The magnetic dipole moment $\mu$ is the only electromagnetic property 
(besides  the 
electric charge) that has been measured experimentally 
\cite{Kotulla:2002cg,Beringer:1900zz}, both for the $\Delta$ and for the $\Omega$. 
We show in Tables \ref{tab:delta_moments} and  \ref{tab:omega_moments} the 
dimensionless quantity $G_{M1}(0)$ which is related to the magnetic dipole via 
$\mu=G_{M1}(0)~e/2M$. In the case of the $\Delta$, our result is consistent with the 
experimental value due to the large uncertainties in the latter; however, the 
experimental value for the $\Omega$ is very accurately measured and our 
results from the Rainbow-Ladder Faddeev equation clearly disagree. 
We  interpret this disagreement to be mainly 
caused by the fact that in the chosen truncation effects of meson dressings of 
the baryon quark core are 
absent.\footnote{In case of the  $\Omega$ one expects effects mostly
from a kaon cloud.} 
The comparison with lattice data is more favorable for the $\Delta$ 
since, presumably, due to the high pion mass in lattice calculations pion cloud 
effects are also suppressed in the lattice calculations.

\begin{table*}
\begin{center}
\renewcommand{\arraystretch}{1.6}
\begin{tabular}{l|cc|ccc|c|} 
~~~~~~~~~~~~~~~~~~~ ~~~~~~~~~~& 
F-\textbf{I} & F-\textbf{II} & DW1 & DW2 & Hyb. & Exp.\\[2mm]
\hline\hline
$M_\Omega (\textnormal{GeV})$ & 1.65 & 1.80 & 1.76~(2) & 1.77~(3) &
1.78~(3) & 1.672
\\
\hline
$<r_{E0}^2> (\textnormal{fm}^2)$ & 0.27 & 0.27 & 0.355~(14) &
0.353~(8) & 0.338~(9) &
\\
\hline
$<r_{E0}^2> M_{\Omega}^2$ & 0.74 & 0.89 & 0.726~(36) & 0.858~(25) &
0.794~(14) &
\\
\hline
$G_{M1}(0)$ & -2.41 & -2.71 & -3.443~(173) & -3.601~(109) & -3.368~(80) &
-3.52~(9) \\ \hline
$G_{E2}(0)$ & 0.54 & 0.75 & 0.959~(41) &  & 0.838~(19) &
 \\ \hline

\end{tabular}
\caption{Comparison of static results for the $\Omega^{-}$. We compare our results 
for models \textbf{I} and \textbf{II} (F-\textbf{I} and F-\textbf{II}, respectively) with a lattice calculation with
dynamical Wilson fermions at $m_\pi=297$~MeV (DW1), $m_\pi=330$~MeV (DW2) and
with a hybrid action at $m_\pi=353$~MeV (Hyb) \cite{Alexandrou:2010jv}. For
$G_{M1}(0)$ we also compare to the experimental value
\cite{Beringer:1900zz}.}\label{tab:omega_moments}
\end{center}
\end{table*}

The last quantity we show is $G_{E2}(0)$, which is related to the magnetic quadrupole 
moment $\mathcal{Q}$ via $\mathcal{Q}=G_{E2}(0)~e/M^2$. In the Breit frame, it can 
be interpreted as a measure of the deformation from sphericity of the electric charge 
distribution and is therefore a quantity of significant interest. The sign of the 
quadrupole moment thus indicates the overall shape of the distribution; our 
calculations agree with lattice  that both the $\Delta$ and $\Omega$ feature an oblate 
($\mathcal{Q}<0$) charge distribution.
At the quantitative level, as with the magnetic dipole, we observe that lattice 
results and ours agree reasonably well for the $\Delta$ (although in this case 
statistical errors for lattice data are very high) and for the Omega the difference 
becomes more marked.  

The shape of the magnetic dipole distribution is indicated by the magnetic octupole 
moment $\mathcal{O}$. Although in the complete calculation presented in 
\cite{FFpaper} we present evidence that $\mathcal{O}$ is likely positive and, therefore, the 
magnetic dipole distribution is prolate, this quantity is numerically very sensitive 
and we cannot provide a truely reliable value at $Q=0$.

\section{Summary}
 
We have presented results for
the electromagnetic moments of the $\Delta$ and $\Omega$ baryons 
calculated from a self-consistent system of Dyson-Schwinger equations for the quark
propagators and the quark-photon vertex as well as three-body Bethe-Salpeter type
equations for the baryon quark core. 
We truncated the system to a single dressed-gluon exchange and used two 
different models for the effective interaction 
to obtain some rough estimate of the model 
dependence of the obtained results. 
These are in reasonable agreement with corresponding lattice data. However, in the 
case of the magnetic moment of the $\Omega$, the only case considered here where
sufficiently precise experimental data is available, the obtained results 
display a qualitatively correct behaviour but are on the quantitative level 
unsatisfactory. 

Such disagreements might be related to the employed Faddeev
approximation ({\it i.e.}, neglecting three-particle irreducible interactions), 
deficiencies in the Rainbow-Ladder approximation and the modeled interaction and/or 
neglecting mesonic dressing effects. To this end we note that for mesonic bound states 
significantly  more sophisticated truncation schemes have been successfully 
employed, see {\it e.g.}, 
Refs.~\cite{Fischer:2007ze,Fischer:2008wy,Chang:2009zb,Fischer:2009jm},
and can be, in principle, also applied to baryons as three-quark bound states.

\section*{Acknowledgements}
We are grateful to the organizers of the {\it Xth Quark Confinement and the Hadron
Spectrum} conference for all their efforts which made this conference possible.\\
We thank Gernot Eichmann, Christian S.\ Fischer and Selym Villalba-Chavez for
helpful discussions.\\
During the course of this work HSA was funded by the Austrian Science Fund, FWF,
under project  P20592-N16
and supported by the Doctoral Program on Hadrons in Vacuum, Nuclei, and Stars 
(FWF DK W1203-N16);
RW acknowledges funding by FWF under project  M1333-N16.
Further support by the  European Union
(HadronPhysics2 project ``Study of strongly-interacting matter'') is acknowledged.

\end{document}